# RTL-Arrow: Hardware-to-Cloud Bridge

**Calvin Deutschbein, Jimmy Ostler**

ckdeutschbein@willamette.edu, jtostler@willamette.edu

## Abstract

Hardware Security at Willamette is a Willamette University affiliated research group studying the hardware-software interface of security critical services. Within our program, we noticed many researchers spent considerable development time learning to understand and manually parse traces-of-execution of hardware designs which are used to identifying whether vulnerabilities or weaknesses arise at the hardware, software, or interface level. We propose the "RTL-Arrow" framework, a framework to compile performant binaries which bridge the hardware/data divide. We translate the outputs of simulated hardware execution, as “value change dumps” into modern data science workflows as cloud-ready “dataframes”, to standardize program verification across the hardware and software levels. We describe our approach, its benefits, and lessons learned from the process of packaging and distributing these libraries for our security research program.

## Biography

*Professor Calvin Deutschbein is a computer security and systems researcher and educator. They completed their Ph.D. in Computer Science at University of North Carolina at Chapel Hill under the direction of Professor Cynthia Sturton. Their research focuses on computer security, especially at the level of hardware design, and in the usage of data mining and design specification to achieve security goals. Prior to joining Willamette, Calvin had years of teaching experience at the University of Chicago, the University of North Carolina at Chapel Hill, and Elon University, the number one ranked US News and World Report institution for undergraduate education. Calvin is especially passionate about expanding the impact of computing education and career and job placements for students. Calvin's research on hardware security has been well received by industry partners, including invited talks for Intel Corporation, the Semiconductor Research Corporation, and Cycuity (formerly Tortuga Logic). Within the research community, they have given invited talks at hardware security-oriented venues such as SEC-RISCV and clean-slate. Their lab group, Hardware Security at Willamette, maintains a variety of open-source hardware security tools and regularly publishes on electrical and computer engineering and computer and data science.*

*Jimmy Ostler is an undergraduate researcher in Mathematics and Computer Science (Engineering) at Willamette University. Jimmy Ostler is the inaugural president of the Willamette Computer Science Student Association and a developmental intern at Cloudflare on C++/Rust interoperability.*

# 1 Introduction

As hardware security researchers, we regard the current paradigm in computing hardware as emerging following the collapse of MOSFET scaling (also called Dennard scaling, for Robert H. Dennard), a sibling of Moore's Law. In brief, CPU Clock Rates increased exponentially from 1975, when the scaling first gained its name, until 2006. In 2006, a "power wall" emerged – chips could continue to become more complex, but if all of their circuitry was active at once, they would melt as heat could not be dissipated quickly enough. This led to a marked change in the design of computer hardware, that has had two broad impacts on our research.

First, one response to the end of MOSFET scaling was the emergence of "dark silicon" - portions of a computer chip that are not currently powered. The most obvious example of this is integer versus floating point arithmetic, which take place on different hardware within CPUs. But in modern devices, there are now a wide ranged of specialized, application-specific instructions that are essentially hardware accelerators for different tasks, ranging from cryptographic units to caches. This is a dramatic increase in hardware complexity, and introduces far greater possibilities for security vulnerabilities to exist and be exploited.

Second, the emergence of GPU-centric computing, especially in data scientific applications, has allowed the scaling of computing performance past the power wall by using increasingly parallelized processes. Essentially, GPUs in 2006 were linear algebra accelerators (and still are, to some degree), and many computational tasks can be completed as vector, matrix, or tensor operations rather than scalar operations performed by CPUs. What this means is that in 2025, any computationally expensive workload should be considered for GPU acceleration and, if at all possible, executed on GPU (or some other ASIC) rather CPU circuitry.

These changes, emergent from underlying physical processes, have worked up the abstraction stack and now see relevance in software in two specific ways. First, the importance of dark silicon for physical reasons has dramatically increased the importance of Electronic Design Automation (EDA, or ECAD) to design increasingly sophisticated and application-specific circuits. For our purposes, the most important EDA tools are Hardware Description Languages (HDLs), which are like programming languages that instead describe integrated circuits at register transfer level (RTL), and simulators, which we use to model the behavior of a hardware system for study. Second, the emergence of GPU acceleration as the dominant paradigm in compute and data analytics tasks has, with it, led to the overwhelming importance of specific tabular or columnar data storage methods in order to leverage to device-specific APIs. That is, EDA has itself become EDA-accelerated, though predominantly through the same mechanism that GPU acceleration is applied to any given task.

However, to our knowledge there is no existing framework that allows study of designs at RTL using existing data scientific methodologies. In our initial work in this area, we have relied on various application-specific technologies – a simulator, an inference engine, and a custom translator between the two – but found this approach had poor performance, parallelized poorly, and had many dependencies, including a Python runtime and Java Virtual Machine.

We propose RTL-Arrow, a Rust language framework that translates simulated traces-of-execution of hardware designs into ASF Arrow's tabular format, which is well-suited to both GPU acceleration and a massive parallelization via cloud technologies should as Hadoop or Spark. We describe our problem statement, our proposed solution, discuss the open-source codebase implementing our solution, and provide an example of a case study in which our codebase improved the performance of a hardware security validation task over open-source silicon.

# 2 Problem Statement

To specifically formulate our problem and our requirements, we will consider it as an iterative improvement to the naive “Myrtha” tool. Myrtha, introduced in our prior work “Test, Build Deploy” is a proof-of-concept container package for cloud-native, CI/CD hardware development (Deutschbein and Stassinopoulos, 2025). Many of limitations we discovered while developing Myrtha motivated the formulation of the RTL-Arrow bridge we propose.

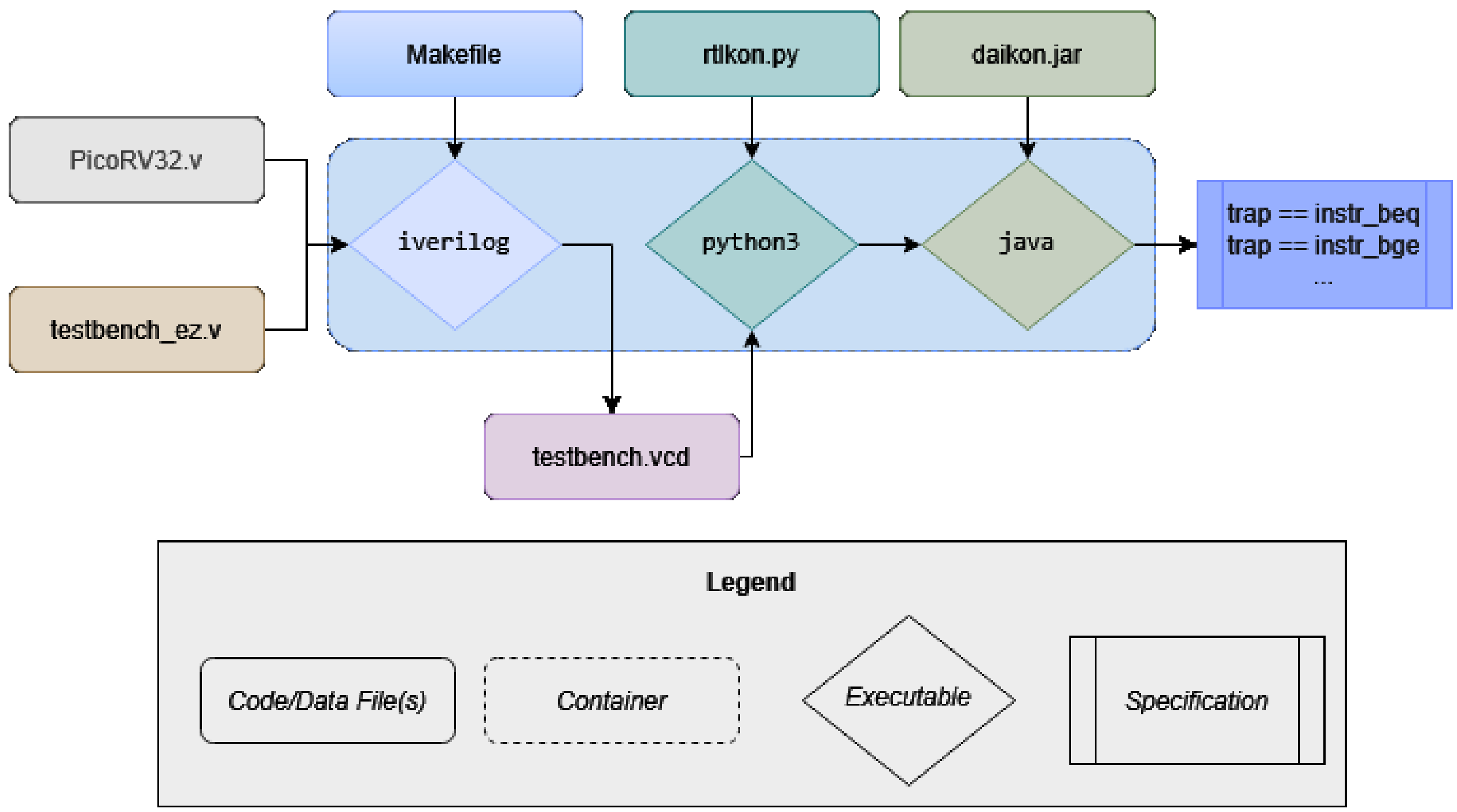

***Figure 1: The “Myrtha” container package for hardware specification.***

We see here the three stages of trace-based hardware validation:

(1) A design is simulated to create a *value change dump* (.vcd) based on some inputs.
    a. In the diagram, we use Icarus Verilog (iverilog) to simulate based on a testbench.
*(2)* The *value change dump* is translated into a *trace of execution*
    a. In the diagram, testbench.vcd is translated via the Python script rtlkon.py
(3) The *trace of execution* is used to infer a *specification*.
    a. In the diagram, inference is performed via the Java program daikon.jar.

In Myrtha, each of these steps are implemented distinctly. The simulator, Icarus Verilog, is written in C and compiled into an executable binary. The translator, “rtlkon.py”, is written in Python and requires a Python runtime. The specification miner, the Daikon Dynamic Invariant Detector, is Java (.jar) application that requires a Java Virtual Machine. Each of these contributes to container size – even Icarus Verilog, which requires the use of a full compiler toolchain and a libc implementation, making extremely lightweight containers difficult.

In our use case, where we are extensively studying hardware designs for difficult-to-detect vulnerabilities, we often use hundreds or more traces to study a single aspect of a design, so any inefficiency within a single container is duplicated many times across our architecture. As it stands, Myrtha is a useful tool for hardware security specification already, but scales poorly. For example, we consider the case of *transient execution CPU vulnerability*.

## 2.1 Transient Execution CPU Vulnerabilities

Transient Execution CPU Vulnerabilities arise from *speculative execution*, a processor optimization wherein, when a pipelined process encounters a branch based on some yet-incomplete computation or memory read, the CPU executes both possible paths pending the result of conditional expression. Branch prediction, and speculative execution are major contributors to modern hardware performance. However, in 2018, the Meltdown (Lipp et al., 2018) and Spectre (Kocker et al., 2019) vulnerabilities were discovered, that made processors utilizing these performance enhancements exceedingly insecure. In brief, an attacker could manipulate branch prediction to speculative issue illegal memory read directives, which would speculatively load privileged data into a software visible cache. When the branch was found to be illegal and terminated, the CPU state would reset, but memory state would not and other attacks could be used to read the sensitive data in memory.

These attacks are elusive because they involve unrecognized disclosure and can be difficult to study for this reason. We think of this in terms of traces – no single trace reveals that a CPU is vulnerable to attack. Instead, multiple traces must be studied, and it must be determined that under no trace does a speculative memory load occur (or some other action) occur.

## 2.2 Scaling and Parallelism

The Myrtha workflow above supports exactly one *trace of execution* at a time, and as a container can be scaled but has scaling performance based on container size and speed. Since we will necessarily use at least one compiled binary for our simulator (whether Icarus Verilog or some other tool, the most likely of which is Verilator), we believe it will be most efficient to generate stand-alone binaries for the remaining stages, containerize these binaries, and then broadly apply the framework to the generation of – and then specification over – many traces. Doing so, we hope to capture specifications that either preclude or detect transient execution CPU vulnerabilities in relatively little real time, and with managably low cloud compute costs. Ideally, we will identify use cases for GPU acceleration for inference tasks versus the CPU-based inference used in earlier tools.

# 3 Proposed Solution

We regard Rust as an industry-standard language for generating performant binaries and have been working to translate our translation and inference stages into Rust for that reason. Alongside other high performance lower-level languages like C++ and Go, Rust has become a popular language for writing cloud-compute lambda functions and has extremely precise memory usage due to borrow checking, which removes the need to include garbage collection in a binary. Rust also supports modern error-handling, which is highly useful to maintain our project across different versions of Verilog and potentially different simulators.

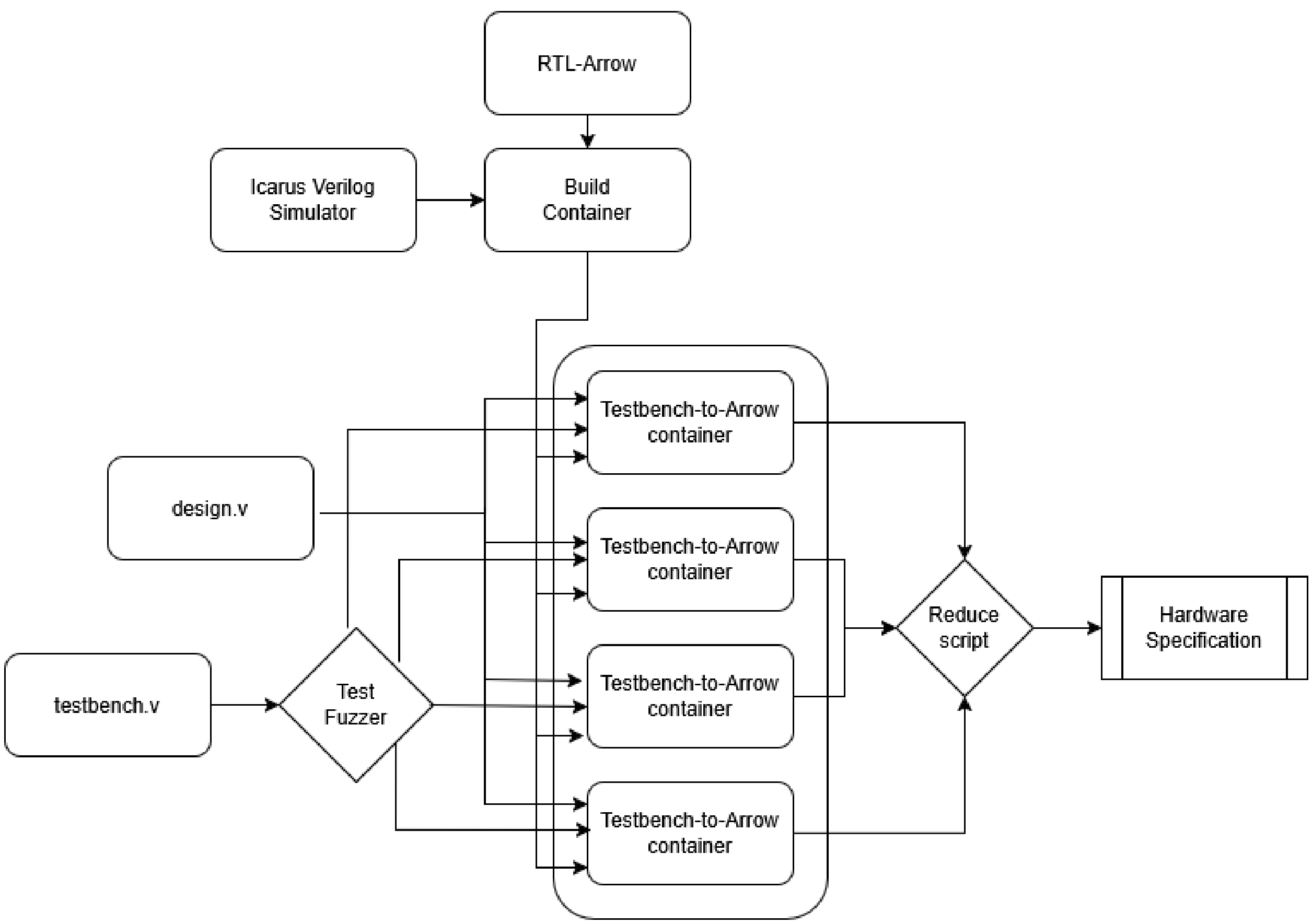

***Figure 2: Proposed Use Case of RTL-Arrow in Cluster***

## 3.1 IEEE 1364 VCD and Daikon

In the "Myrtha" tool, both *value change dumps* and *traces of execution* were stored in an ASCII-based format – the IEEE 1364 VCD format for value change dumps, and the traces in a two-file format specific to Daikon, a dynamic invariant detector which extracts trace properties from simulated traces – and can be used to automatically generate security properties. Both formats utilize a header, which enumerates the registers (and wires) of a design.

The *value change dump* then logs any changes to register values from the starting, uninitialized value. By contrast, the Daikon trace enumerates the value of every register at every clock tick – which is modelled as a program point. Internally, Daikon models the hardware as a program, and each clock tick of the processor is modelled as a method or function call with its own name space.

When initially released in 2004 (Ernst, 2007), and even when initially applied to hardware in 2017, the computing ecosystem was entirely different, and Daikon's trace format was unremarkable. This no longer the case.

While Python has been the most popular language year-on-year since 2017, and boasts a wide range of inference and analytics libraries, in 2004 there were no existing interchange formats – even Hadoop had not yet been released – and even in 2017 GPU acceleration inference was in its infancy and rarely applied to domains outside of vision.

Now, in 2025, there are well-supported and generalized cloud and GPU capability data interchange formats that suitable any number of tasks, including program and hardware validation. As recently as the past 12 months in the past year, we have seen two major developments that impact the relevance of our work. The most popular tabular format in Python – **pandas –** released a new version that is now ASF

Arrow-backed. Separately, the ubiquitous cloud compute technology ASF Spark has released a Rust kernel to support Rust and Python user-defined functions. These design decisions were made with good reason and should be applied to hardware validation as well.

## 3.2 Arrow

Internally, we can eliminate both ASCII-formatted write-to-disk memory costs by utilizing the Arrow format:

> "universal columnar format and multi-language toolbox for fast data interchange and in-memory analytics" (Apache Arrow, 2025)

We can do so even with minimal modifications to the existing simulator by piping the data from the simulator to our encoding process. The capturing application need only then maintain minimal working state – current values of all registers – and then write the values to a dataframe at each time point.

Arrow helpfully has first-class support in a variety of languages, ranging from interpreted scripting languages like R and Python to systems languages like C/C++ and Rust.

Once a *trace of execution* has been placed in an Arrow format, it can be written to disk in a machine-readable binary format, either the Arrow format itself for zero-copy reads or the compressed Parquet format used for storage. In general, we expect to write to Parquet anticipating revisiting trace data numerous times during the security research process.

## 3.3 Polars

To meet our requirements for a compact binary, we wanted to use a compiled language with no garbage collector, eliminating a variety options, perhaps most obviously Go. Fortunately, Rust has excellent library support for working with Arrow, much like C/C++ which have Arrow libraries, but Rust additionally has a native, Arrow-backed dataframe library – Polars (to the best of our knowledge, C++ has both dataframe support and first-class Arrow support, but does not have an Arrow-backed dataframe library).

We plan to use Polars to translate our *value change dump* files into Arrow dataframes, which can be accepted by a variety of modern inference engines implementing many of the clustering algorithms used within Daikon. We propose a performant RTL-to-Arrow stage combining a high-performance Verilog compiler with a high-performance Arrow binary, each in a compiled language with no garbage collection.

### 3.3.1 LazyFrames

A core feature of Polars is the LazyFrame – a dataframe that delays evaluations of transforms until being queried. Within a data science technology stack, this can lead to considerably improved performance by optimizing the sequences of transforms and operations. For our purposes, we have no real benefit to using LazyFrames as we simply treat Polars as an Arrow API.

We do, however, intend to take advantage of LazyFrames within our analytics frameworks after completing translation, and still regard the LazyFrame as a core benefit of our language selection for that reason. We motivate the benefits with an example.

Suppose we are considering a design with 1000 registers, of which 200 each are distributed over an ALU, FPU, MMU, debug interface (DBG), and interrupt handler (IRQ). We may wish to validate the memory calculations of this design to ensure that no arithmetic errors contribute to illegal reads. To do so, we likely will complete two tasks:

1. Functional validation of the ALU's arithmetic capabilities
2. Information flow analysis from the FPU to MMU to ensure float values are not dereferenced.

In both cases, we will issue queries not over the entire dataframe, but over subsets of the dataframe, and compose queries across multiple subsets (especially in the case of two).

By executing lazily, these queries may be fully specified and then reordered by the underlying library into a performance optimized sequence. This can avoid redundant or unnecessary calculations, and exploit performance optimizations not exposed to the end user. Rather than learn the intricacies of optimizing dataframe queries, we can use LazyFrames as a best effort optimizer, and only manually optimize in the case of specific performance requirements for specific queries.

Lazy evaluation is non-trivial to implement in languages without built-in support (and in fact the Python implementation is simply a Rust API) and our impressive of the dramatic rise in the popularity of Polars is consistent with this implementation detail driving major performance improvements across a range of domains, including, we hope, hardware validation.

# 4 Design Decisions

## 4.1 Nullable types

One of the great benefits of working with Rust/Polars/Arrow triple is the treatment of *nullable integers,* which are a common and important case when studying hardware traces. Hardware signals generally are treated as binary, but IEEE 1364 they are more properly quartrenary values which can be represented at bit level as ‘0’, ‘1’, ‘x’, or ‘z’. In particular, ‘x’ values are extremely common to represent hardware signals which have not yet been initialized to values. In a security context these signals are often important due to many security agreements around design states before and across the time point when a hardware device comes out of reset.

In prior work (vcd2df, 2025 and Deutschbein, 2025), it was difficult to represent these values as many data-oriented scripting languages lack graceful support for nullable integers of fixed width. In a 32 bit design, for example, there are many registers which are 32 bits in size but many represent greater than $2^{32}$ distinct values, but far fewer than $2^{64}$. Neither Python nor R, for example, can store these values efficiently. However, Rust supports an “Option” type:

> “Type Option represents an optional value: every Option is either Some and contains a value, or None, and does not.”

This is precisely what our implementation demands. Either register either is initialized and holds a numerical value, or is not and does not.

Separately, Arrow supports nullable data at dataframe level by maintaining a *validity bitmask*, stored adjacent to the dataframe in memory (Tustvold, and Alamb, 2022).This *validity bitmask* allows storing options with the minimal cost of a single additional bit per value – effectively allowing 32 bit option storage within only 33 bits.

And finally, the Polars dataframe library helpful stores Rust options directly as nullable data within Arrow dataframes, posing no additional implementation overhead to us as designers or clients of the software.

## 4.2 Columnar vs. Record-Based

A natural question about using a dataframe format is the trade-offs between columnar formats, which represent some observations as a matrix or two-dimensional array, or record-based formats like JSON which represent observations as leaf nodes in a tree.

One common benefit to dataframe encoding is compression, which importantly is not a benefit in our case. When modelling hardware designs, most registers are unchanged cycle-to-cycle, so we experience massive data duplication and frequently binary dataframe formats have larger storage footprints than text-based record formats. And we do consider memory in our case, especially for larger designs or longer test suites.

That said, the overwhelming benefit of dataframe support for queries cannot be understated. Prior efforts in hardware validation often had to develop sophisticated, custom parsers to infer relatively simple relationships between registers, such as clustering algorithms or binary relations. In our experience, any query required extensive implementation time to interface directly with the underlying trace data.

Dataframes offer high performance due to vectorization and parallelization to most forms of queries, and often support predictable, constant-time memory access. We have found dataframe storage to be accelerant for hardware security research within our group, and believe the storage costs are justified by the increased development speed.

That said, it merits further investigation to determine what, if any, benefits can be derived by introducing both utilities for translation to record-based types, and the usage of a binary record-based storage medium such as bincode (bicode-org, 2025), especially over larger designs.

# 5 Results

To assess the comparative performance of RTL-Arrow versus the state-of-the-art, we will compare our RTL-Arrow implementation against libraries in Python and to translate VCD files to Arrow formats. In a few cases, we will have to modify the packages slightly, but the points of comparison are helpfully open-source and we are able to make minor changes.

To do so, we implemented a crate in Rust and performed a simple example, based on a use case involving a map of information flow detection across 181 *value change dump* files. In both cases, we output a series of pairwise information flow events as JSON.

## 5.1 Compilation Cost

To provide a fair comparison to interpreted languages, we believe the closest point of comparison will be to use a *build container* to configure a *cluster container*. The best point of comparison, we believe, is comparing the container build time for either (1) installing the relevant runtime on the cluster container for an interpreted technology or (2) installing the compiler toolchain on the build container, compiling, and copying to the cluster container for a compiled technology.

We do not necessarily expect RTL-Arrow to be particularly competitive at this stage. Both the Rust compiler installation (rustup) and compiling with the Polars crate are fairly involved.

In practice, we found that through the use a *build container* we were able to avoid the costly Rust installation using a community-provided pre-built image, but we still experienced costly installation of Arrow and Polars, which we used to interface with Arrow.

Our first effort to instantiate a Python container completed in 74.795 seconds on our reference system. By contrast, the Rust variant required 291.691 seconds. Testing locally, we found Rust compilation took 19.600 seconds after installing dependencies.

We did not see significant differences between the use of Podman or Docker and saw only a spread of a few seconds across repeated trials. In both cases, we began with a pre-built image maintained by the language foundation, installed all language packages through the language package manager ("pip" for Python and "cargo" for Rust) and used a "git" binary to retrieve the files for the tests from a reference repository (cd-public, 2025).

## 5.2 Container Size

Our Python image, which contained the Python based images and the minimal Python "vcd2df" library, required 1.76 GB of storage, of which approximately 1.11 GB was the base image and the remaining was "pandas" dataframe library and its dependencies.

Our Rust build image was 2.59 GB in size, of which approximately 1.79 was the base image and the remainder was Arrow and Polars dependencies. Importantly, we note this also must contain a linker, for which we used “gcc”.

Our Rust cluster image was 778 MB in size, which we expect to overwhelmingly be Arrow, as this memory footprint is quite similar to the additional size needed for a Python image to interface with Arrow, and the cluster image should contain little else beyond a minimized OS (for which we used Debian).

### 5.3 Run speed

Running locally, we found Python completed the security characterization task in 28.468 seconds. Repeated runs fell within a second.

By contrast, our Rust implementation, which we had regarded as almost fully unoptimized, ran in 2.217 seconds. Repeated runs fell within a tenth of a second.

We found this performance unexpectedly high. Having specifically worked in Python for this research direction for close to a decade, we have few remaining optimizations to accelerate our Python code beyond deploying a cluster. By contrast, we do not regard ourselves as beyond entry-level Rust programmers and were concerned a lack of familiarity with the language may incur heavy costs due to unnecessary borrows or poor choice of types. In point of fact, we appeared to approach the I/O bound without optimizations, falling with an order of magnitude of “wc” (at .285 seconds).

# 6 Conclusion

Despite relatively high costs for development and compilation, the usage of a Rust binary markedly reduced our memory footprint on highly parallelizable process for hardware security research. We also found several unexpected benefits – first class support for an option type in Rust greatly improved our ability to interface with nullable integers, a natural implementation for potentially uninitialized hardware signals, and the Rust string APIs greatly simplified the process of parsing an relatively undocumented file type.

We encourage similar organizations preparing to scale to larger data science applications to consider moving repeated and parallelizable routines into the Rust language, and look forward to how it can accelerate our own research efforts in the coming years.

We found less than half the memory footprint and over ten times the performance for fairly minimal development costs.